\pdfoutput=1 
\documentclass[%
 aip,
 amsmath,amssymb,
 reprint,%
]{revtex4-1}
\usepackage{tabularx}
\usepackage[labelformat=simple]{subcaption}

\usepackage{graphicx}
\usepackage{dcolumn}
\usepackage{bm}
\usepackage{siunitx}
\usepackage[utf8]{inputenc}
\usepackage[T1]{fontenc}
\usepackage{mathptmx}
\usepackage{etoolbox}
\usepackage[justification=justified]{caption}
\makeatletter
\def\@email#1#2{%
 \endgroup
 \patchcmd{\titleblock@produce}
  {\frontmatter@RRAPformat}
  {\frontmatter@RRAPformat{\produce@RRAP{*#1\href{mailto:#2}{#2}}}\frontmatter@RRAPformat}
  {}{}
}%
\makeatother
\begin{document}

\title[]{X-ray diffraction with micrometer spatial resolution for highly absorbing samples}

\author{P.~Chakrabarti}
\affiliation{Physics Department, University of Siegen, 57072 Siegen, Germany}
\affiliation{Center for X-ray and Nano Science CXNS, Deutsches Elektronen-Synchrotron DESY, 22607 Hamburg, Germany}
\email{prerana.chakrabarti@uni-siegen.de}

\author{A.~Wildeis}
\author{M.~Hartmann}
\author{R.~Brandt}
\affiliation{Mechanical Engineering Department,  University of Siegen, 57076 Siegen, Germany}

\author{R.~D\"ohrmann}
\author{G.~Fevola}
\affiliation{Center for X-ray and Nano Science CXNS, Deutsches Elektronen-Synchrotron DESY, 22607 Hamburg, Germany}
\author{C.~Ossig}
\affiliation{Center for X-ray and Nano Science CXNS, Deutsches Elektronen-Synchrotron DESY, 22607 Hamburg, Germany}
\affiliation{Physics Department, University of Hamburg, 22761 Hamburg, Germany}
\author{M.~E.~Stuckelberger}
\affiliation{Center for X-ray and Nano Science CXNS, Deutsches Elektronen-Synchrotron DESY, 22607 Hamburg, Germany}

\author{J.~Garrevoet}
\author{K.~V.~Falch}
\author{V.~Galbierz}
\author{G.~Falkenberg}
\affiliation{Deutsches Elektronen-Synchrotron DESY, 22607 Hamburg, Germany} 

\author{P.~Modregger}
\affiliation{Physics Department, University of Siegen, 57072 Siegen, Germany}
\affiliation{Center for X-ray and Nano Science CXNS, Deutsches Elektronen-Synchrotron DESY, 22607 Hamburg, Germany}

\date{\today}

\begin{abstract}
X-ray diffraction with high spatial resolution is commonly used to characterize (poly-)crystalline samples with, for example, respect to local strain, residual stress, grain boundaries and texture. However, the investigation of highly absorbing samples or the simultaneous assessment of high-$Z$ materials by X-ray fluorescence have been limited due to the utilisation of low photon energies. Here, we report on a goniometer-based setup implemented at the P06 beamline of PETRA III that allows for micrometer spatial resolution with a photon energy of 35~keV and above. A highly focused beam was achieved by using compound refractive lenses and high precision sample manipulation was enabled by a goniometer that allows for up to 5D scans (3 rotations \& 2 translations). As experimental examples, we demonstrate the determination of local strain variations in martensitic steel samples with micrometer spatial resolution as well as the simultaneous elemental distribution for high-$Z$ materials in a thin film solar cell. Our proposed approach allows users from the materials science community to determine micro-structural properties even in highly absorbing samples.
\end{abstract}

\maketitle

\section{\label{sec:Introduction}Introduction}

Local deformations of (poly-)crystalline structures, such as strain, tilt or grain boundaries, can have a profound impact on the performance of functional and electronic materials. Two examples are the conversion efficiency of solar cells or the fatigue strength in spring steels. High resolution micro X-ray diffraction ($\mu$-HRXRD) utilizes focused X-ray beams to  study local defects on the nanoscale. 

Examples for $\mu$-HRXRD include the study of local strains and grain boundaries in functional thin films with a photon energy of 8.9~keV and a spot size of $(100\times 100)\ \mathrm{nm}^2$ at the ID01 beamline of the European Synchrotron Facility~\cite{microdiffraction:thinfilm}, a similar characterization of thin film photovoltaic cells with similar experimental parameters at the Hard X-Ray Nanoprobe (HXN) beamline 3-ID of the National Synchrotron Light Source II ~\cite{strain:photovoltaic,ulvestad-2019-jsr}. Other examples are the determination of local micro-structure of martensite-retained austenite steel with a photon energy of 12~keV and a spot size of $(4\times 1)\,\mu \mathrm{m}^2$ at the 2-ID-D undulator beamline of the Advanced Photon Source~\cite{microdiffraction:cai} or the renovated high-pressure XRD setup at the BL10XU beamline of SPring-8~\cite{SPring-8:renovation} utilizing a photon energy of 30~keV and a spot size of $(1\times 1)\,\mu \mathrm{m}^2$. 

Investigations with high spatial resolution of highly absorbing samples rely on high photon energies and, simultaneously, small spot sizes. Here, we report on a novel goniometer-based setup implemented at the P06 beamline of PETRA III that enables $\mu$-HRXRD with a few micrometer spatial resolution at a photon energy of 35~keV. In addition, the high photon energy allows for the simultaneous acquisition of the X-ray fluorescence (XRF) signal of elements up to iodine.

\section{\label{sec:Experiment}Experimental setup}

\begin{figure}[htb]
\centering
\captionsetup{justification=raggedright,singlelinecheck = false}
\includegraphics[width=0.48\textwidth]{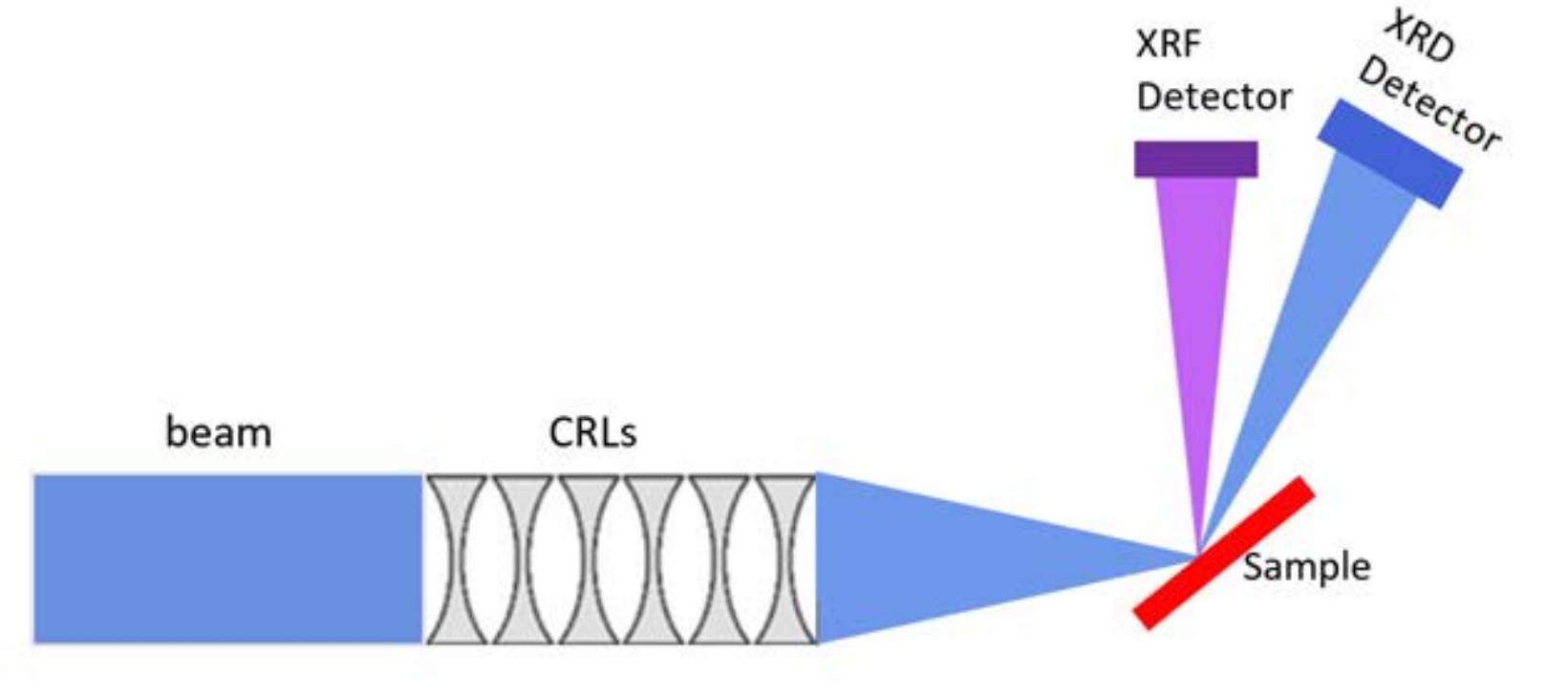}\\
(a)\\
\includegraphics[width=0.38\textwidth]{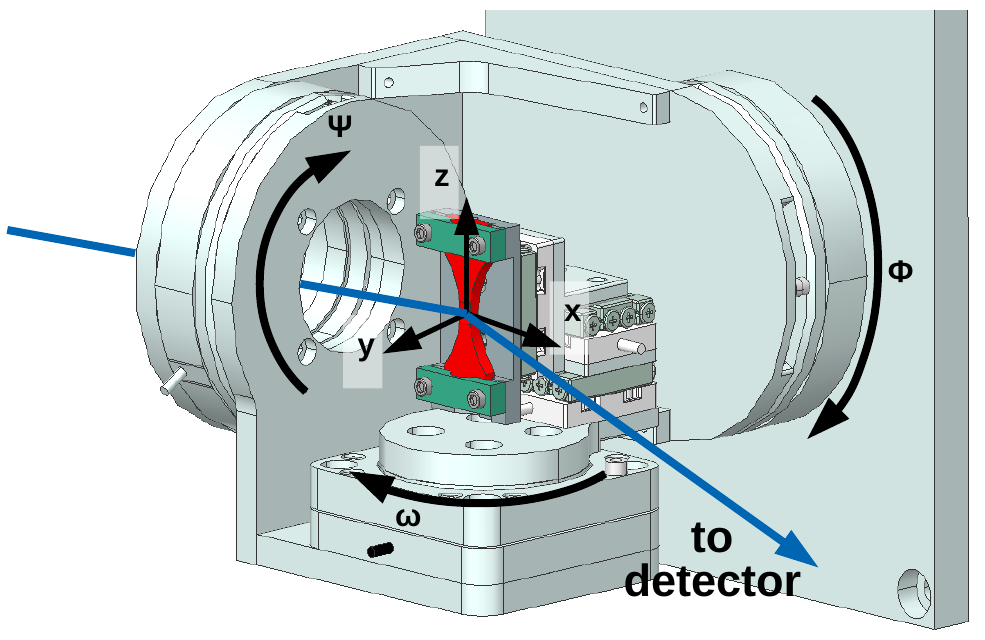}\\
(b)%
\caption{\label{fig:setup}(a) Setup geometry (top view). (b) Technical drawing of the goniometer.}
\end{figure}

All measurements presented here were obtained at the microhutch of beamline P06 of PETRA III at DESY, Hamburg~\cite{Falkenberg:2020}. The incident photon energy of 35~keV was selected by a double crystal monochromator. Compound refractive lenses (CRLs) were used to focus the beam to a spot size of $(2.0\times 1.2)\,\mu \mathrm{m}^2$, which was measured by knife-edge scans of crossed gold wires. Focal distance of the CRLs was 660~mm measured from the pinhole exit (0.4~mm diameter) of the $N_2$-rinsed CRL box, which  lead to a longitudinal spot size of a few millimeters and a photon flux above $10^9$ at the focal position~\cite{Falkenberg:2020}. 

A 6-axes goniometer (SmarAct) fixed on a kinematic mount was used for sample manipulation (see Fig.~\ref{fig:setup}b). It featured 3 rotation circles with nominal angular resolutions below 0.1~$\mu$rad. Additional x, y and z translation stages on top of the last rotation enabled horizontal alignment along the beam (x) and scanning of the sample (y $\&$ z) with fixed rotation angles and nominal position accuracy of 1~nm. The goniometer provides up to 5D scans for characterization of reciprocal space as a function of sample position. The X-ray diffraction signal was obtained with a GaAs Lambda 2M detector (XSPECTRUM) with 55~$\mu$m pixel size, 24-bit range and 1~ms readout frequency approximately 1~m downstream of the sample (Fig.~\ref{fig:setup}a). The X-ray fluorescence signal was obtained by a Vortex SDD (Hitachi High-Tech) located at 90 degrees from the incident beam. The setup allowed for photon energies up to 42~keV.

The cross-point of the rotation axes defines the desired scan position on the sample. Alignment of the cross-point into the beam focus was achieved by a two step procedure: first, an optical microscope was used to move the tip of a tomographic pin into the cross-point and, second, the XRF signal of the pin was used to move the entire goniometer and, therefore, the cross-point into the beam focus.

\section{\label{sec:Results}Results}

Two types of samples were scanned. Martensitic steel as an example for a powder-like structure as well as a thin film solar cell as an example for a single crystal structure.

Martensite (i.e., $\alpha '$-iron) is a meta-stable phase of carbon steel, which is distinguished by its micro- and nanoscopic lamellar structure~\cite{Krauss1999} resulting in a powder-like material system in the given context. Martensitic steels provide ultimate tensile strengths beyond 2~GPa and exhibit an excellent resilience in the high cycle fatigue regime, which renders them as the material of choice for suspension springs in vehicles~\cite{Tump2016}. Martensite is produced by first heating specific iron carbon alloys to temperatures around 1000~$^{\circ}$C, where austenite (i.e., $\gamma$-iron) with typical grain sizes of a few 10~nm is formed. Subsequently, the material is quenched, which prevents a diffusion of carbon atoms out of the crystal structure and allows the highly strained martensitic phase to form. During this process, martensite grows in a lamella-like structure in distinct directions within each prior austenite grain. The fatigue strength of martensitic steels can be further improved by introducing compressive residual stress via shot peening~\cite{Eleiche2001}, which constitutes the inducement of plastic deformation via striking the material surface with a blasting medium.

Here, we have used martensitic steel of type 54SiCr6, which was supplied in the form of wires with a diameter of 12 mm. Heat treatment consisted of austenization in vacuum at 1080~$^{\circ}$C for 100 min and gas quenching using compressed nitrogen, followed by tempering at 400~$^{\circ}$C for 1~h in an inert Ar atmosphere. Flat shallow notched specimens were then manufactured by means of wire erosion. Subsequent shot peening was performed by using cast iron steel-shot with an average size of 0.4~mm at 1.5~bar. Shot peening intensities are commonly quantified by the Almen strip test, which involves the exposure of standardized steel strips identical to the considered shot peening settings~\cite{Totten2002}. The arc height of the bending of the steel strips is then used to quantify the shot peening intensity. Here, shot peening intensities were 0.16~mm or 0.3~mm using one-sided peened steel strips.

\begin{figure}[htb]
    \centering
    \captionsetup{justification=raggedright,singlelinecheck = false}
    \includegraphics[width=0.4\textwidth]{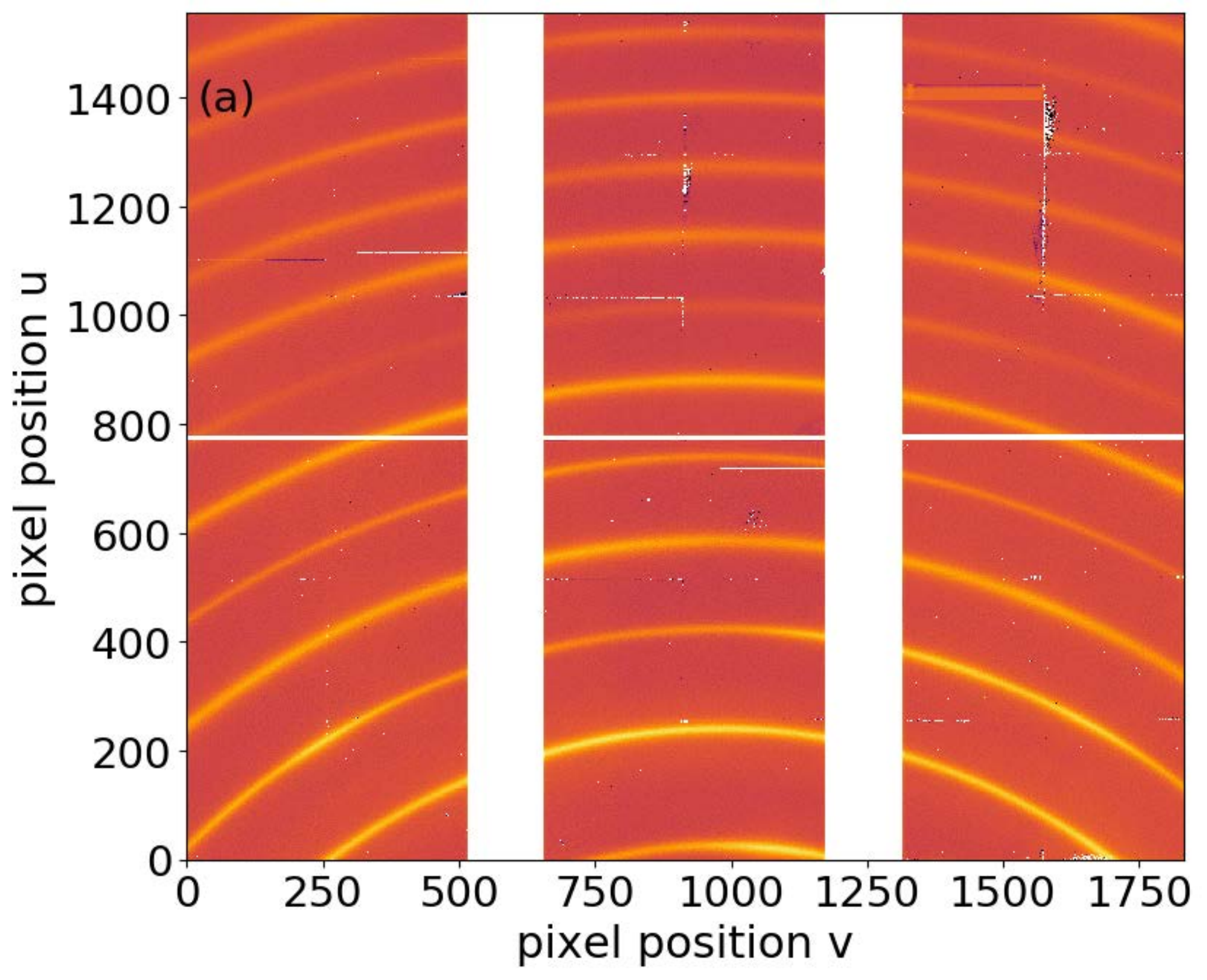}\\
    \includegraphics[width=0.4\textwidth]{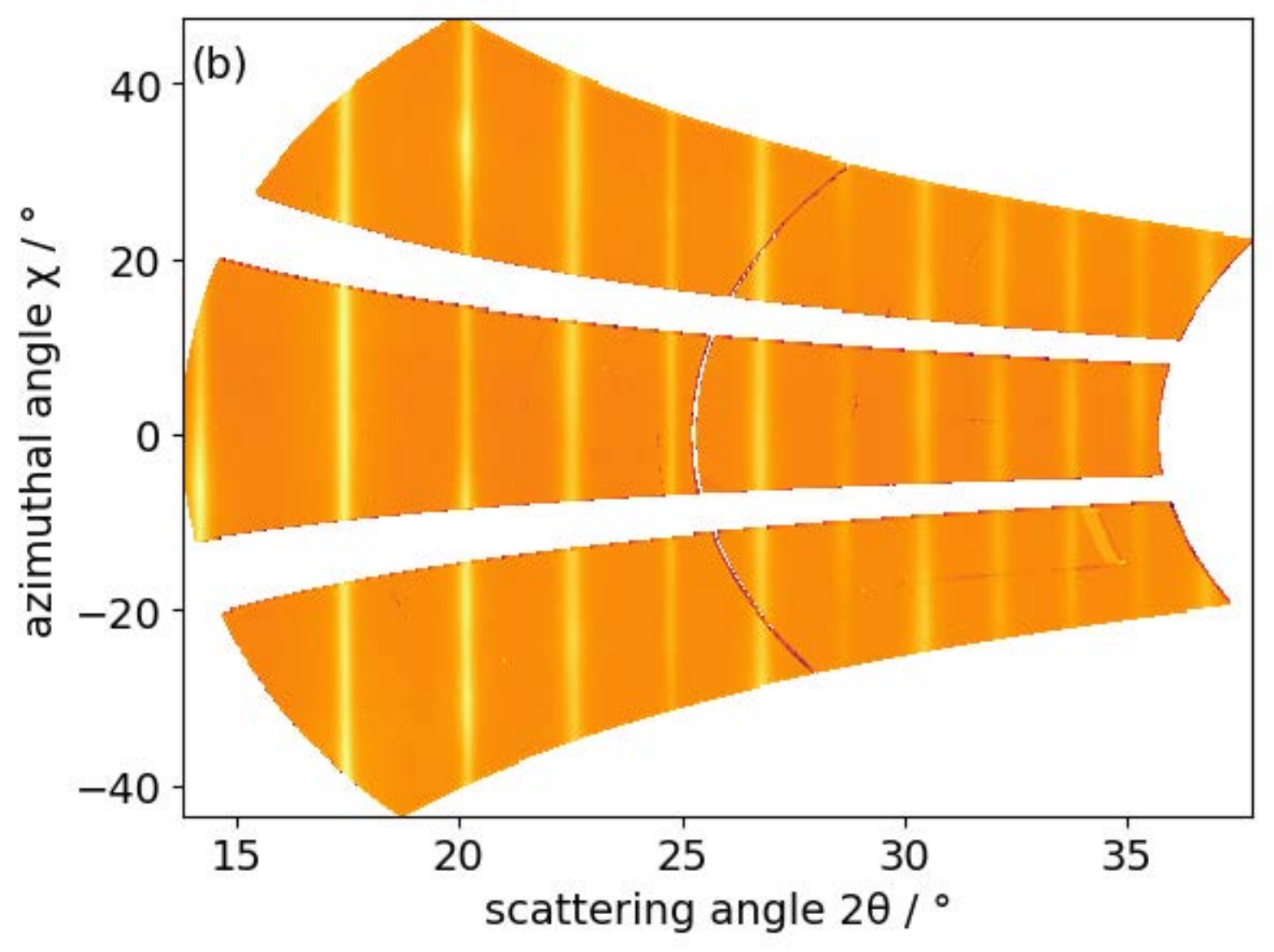}\\
    \includegraphics[width=0.48\textwidth]{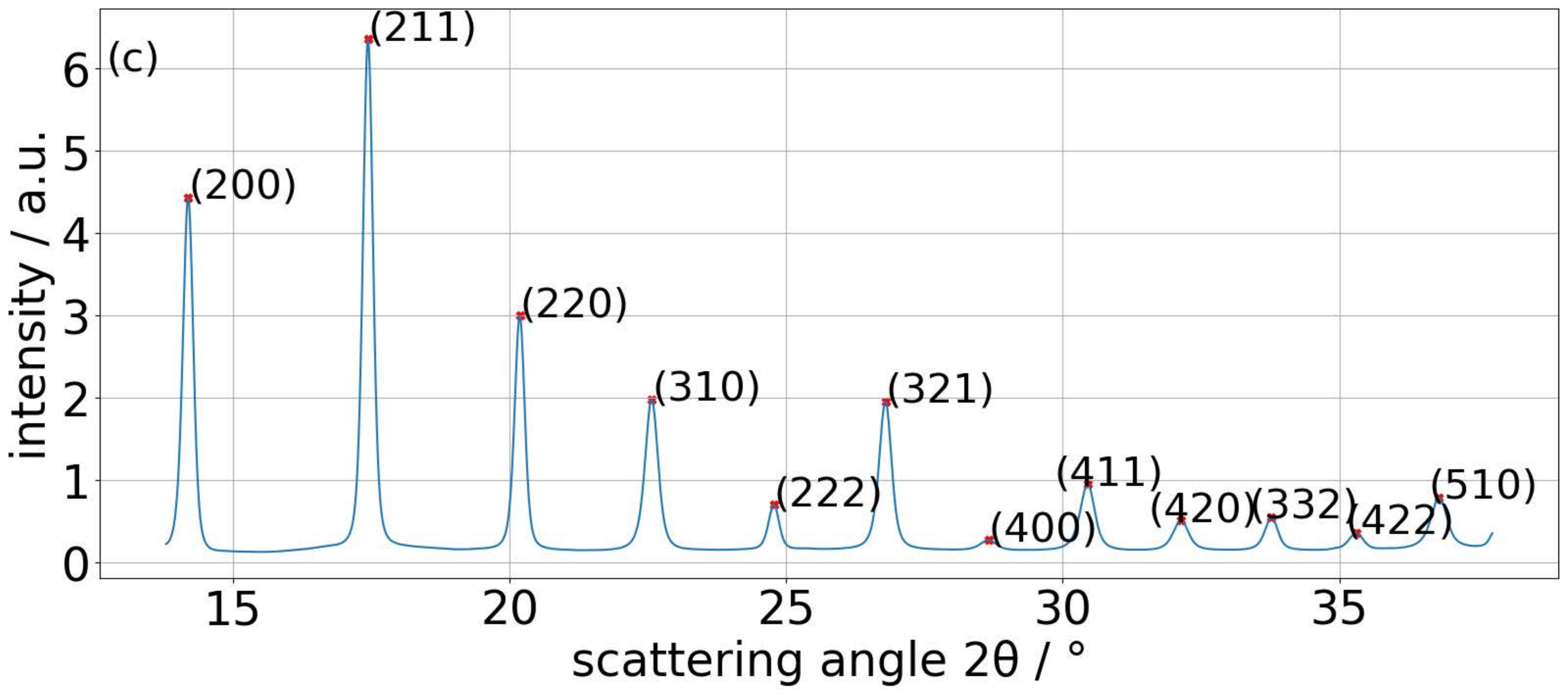}\\
    \caption{(a) Diffraction pattern of a martensitic steel sample summed over 2601 translation scan points. (b) Caking of the diffraction pattern in (a). (c) Final azimuthal integration of (b) yielding the $2\theta$ positions of martensitic diffraction peaks.}
    \label{fig:azimuthal}
\end{figure}


The cross section of the martensite samples was quadratic, where the incident X-ray beam entered the sample through the front and exited through a side. The photon energy of 35~keV implied an absorption length of 250~$\mu$m. The martensite samples were laterally scanned in y- and z-direction with 51 by 51~steps of 2~$\mu$m each and an exposure time between 0.2~s and 2~s. 

In order to translate pixel positions on the detector into $2\theta$-values, we calibrated the setup geometry using the calibration routine provided by PyFAI~\cite{pyFAI:giannis, pyFAI:jerome} and lanthanium hexaboride (LaB$_6$) as a diffraction standard. For automatic peak detection all the diffraction patterns were summed up (Fig.~\ref{fig:azimuthal}a). Single diffraction patterns were transformed from Cartesian coordinates ($u$, $v$) to polar coordinates (2$\theta$, $\chi$) by so-called caking (or regrouping)~\cite{pyFAI:jerome} and an example result is shown in Fig.~\ref{fig:azimuthal}(b). Subsequent vertical (i.e., azimuthal) integration provided the diffraction signal over 2$\theta$. Fig.~\ref{fig:azimuthal}(c) shows that the first 12 martensitic diffraction peaks (i.e., from (200) to (510)) were detected.

For the determination of angular position $\mu$, angular width $\sigma$, height $H$ and background $C$ of all occurring diffraction peaks, the azimuthal integrated intensities were fitted to a Gaussian distribution~\cite{gaussian}, given by
\begin{equation} \label{eq:gauss}
f(2\theta) = C+H \exp\left( -\frac{1}{2}\left(\frac{2\theta-\mu}{\sigma}\right)^{\!2}\,\right).
\end{equation}
Local strain for the reflection $hkl$, $\epsilon_{hkl}$ is given by the differential Bragg equation~\cite{differentialbragg}
\begin{equation} \label{eq:strain}
\epsilon_{hkl}=\frac{d - d_{0}}{d_{0}} = - \Delta\theta_{hkl} / \tan \theta_{hkl}
\end{equation}
with $d$ and $d_{0}$, the net plane spacing for strained and unstrained materials, respectively~\cite{sinesquared}. While martensite crystallizes in a BCT lattice and the $a/c$ ratio depends for example on carbon content~\cite{Sherby2008}, here we assume that $a/c = 1$, which yields a BCC structure. Further, the lattice constant of unstrained martensitic steel was taken as $a  = 2.866$~\AA~\cite{Kim2014}. However, the sampling of the LaB$_6$ diffraction patterns turned out to be insufficient for the determination of absolute strains and, thus, only relative strains will be reported in the following.

\begin{figure}[htbp]
\centering
\captionsetup{justification=raggedright,singlelinecheck = false}
\includegraphics[width=0.45\textwidth]{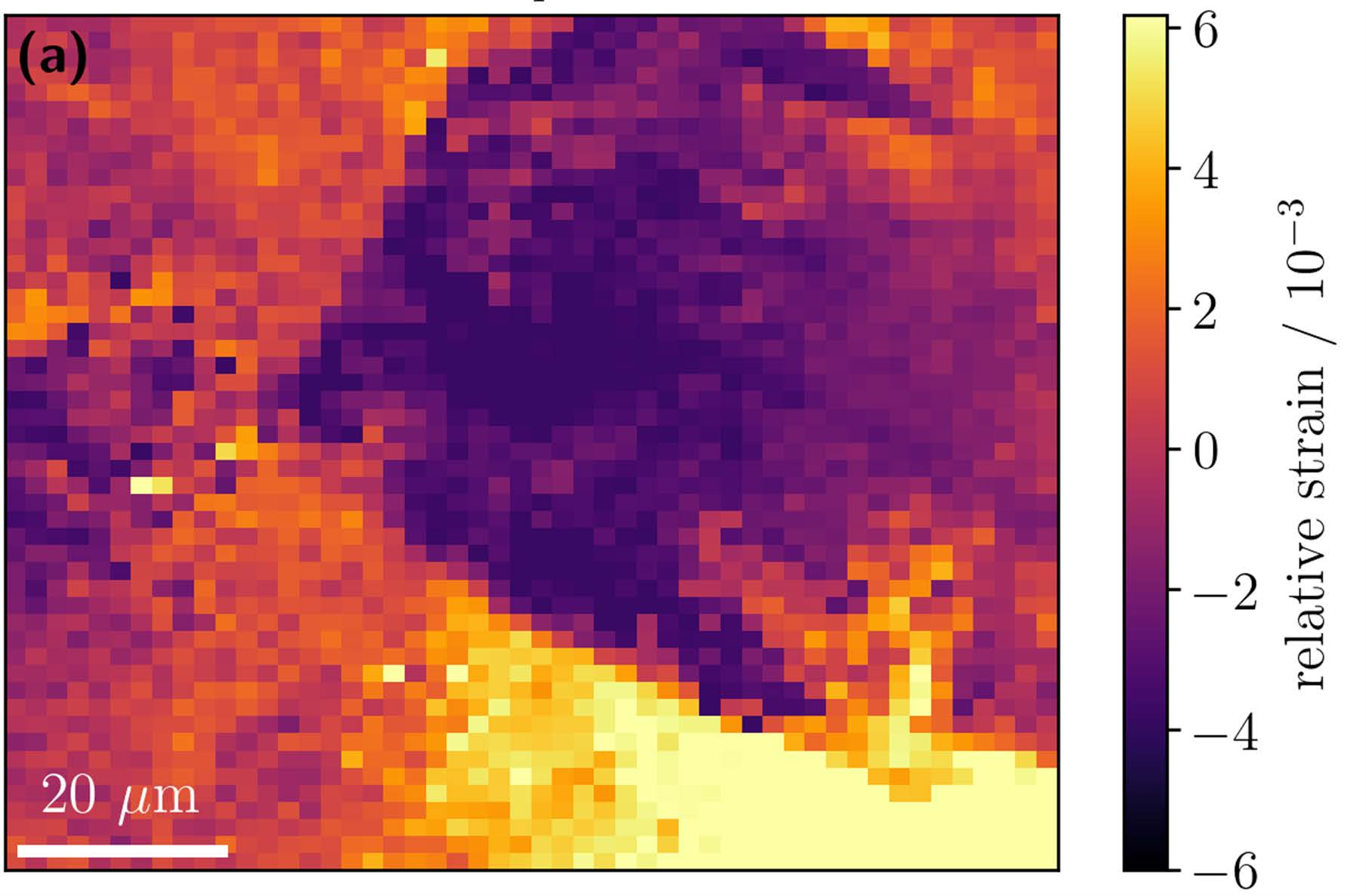}\\
\includegraphics[width=0.45\textwidth]{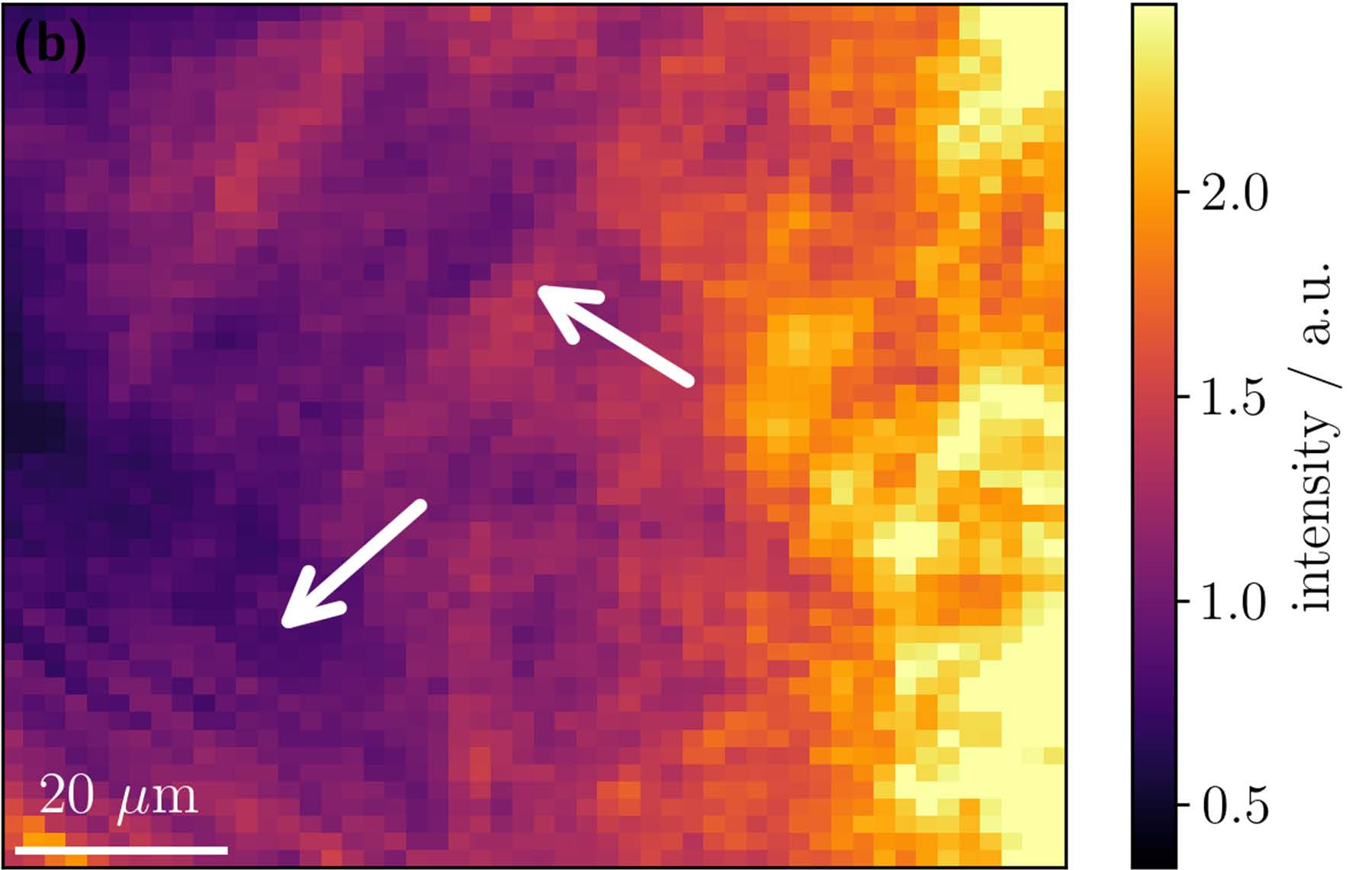}\\
\caption{\label{fig:data analysis_1}(a) Strain $\epsilon_{200}$ as a function of sample according to eq.~(\ref{eq:strain}). (b) Intensity ($C+H$ in eq.~\ref{eq:gauss}) of the (310) reflection showing lamella-like martensite structure. Although the images show the same area of the sample, the visible information relating to the two different reflections is complementary and originates from different structures.}
\end{figure}

Fig.~\ref{fig:data analysis_1}(a) shows the relative strain in the [200] direction, $\epsilon_{200}$ as a function of position on the martensitic steel sample. The boundaries of one prior austenite grain is clearly visible. Fig.~\ref{fig:data analysis_1}(b) shows lamella-like structure of martensite in the intensity of the (310) reflection. The apparent dissimilarity between the images can be readily explained by the following. Due to the relatively large absorption length of 250~$\mu$m, martensitic structures of about 10 to 20 randomly oriented prior austenite grains may contribute to the observable diffraction. Thus, the different reflections show different structures in Fig.~\ref{fig:data analysis_1}(a) and (b). 

\begin{figure}[htb]
\centering
\captionsetup{justification=raggedright,singlelinecheck = false}
\includegraphics[width=0.45\textwidth]{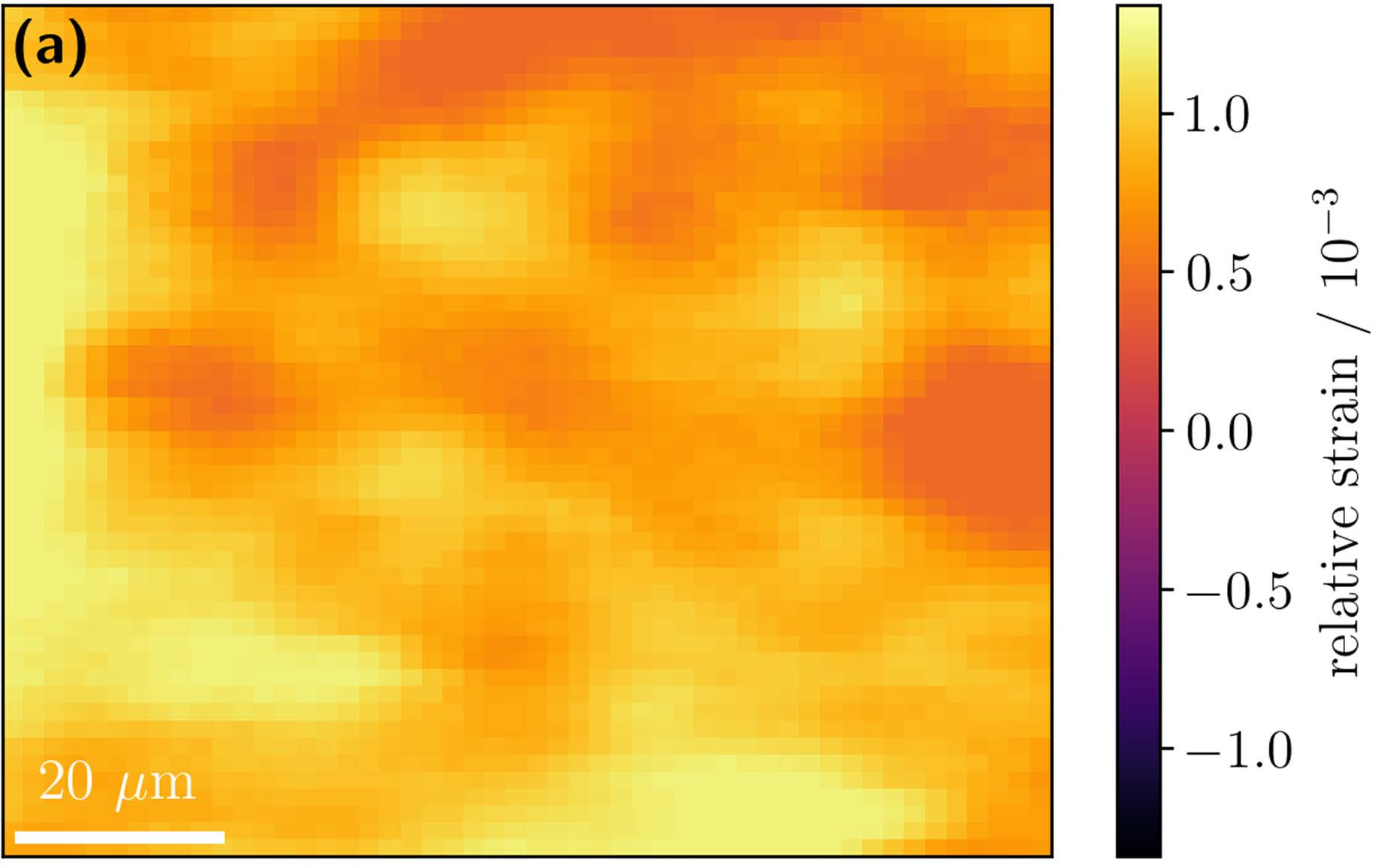}\\
\includegraphics[width=0.45\textwidth]{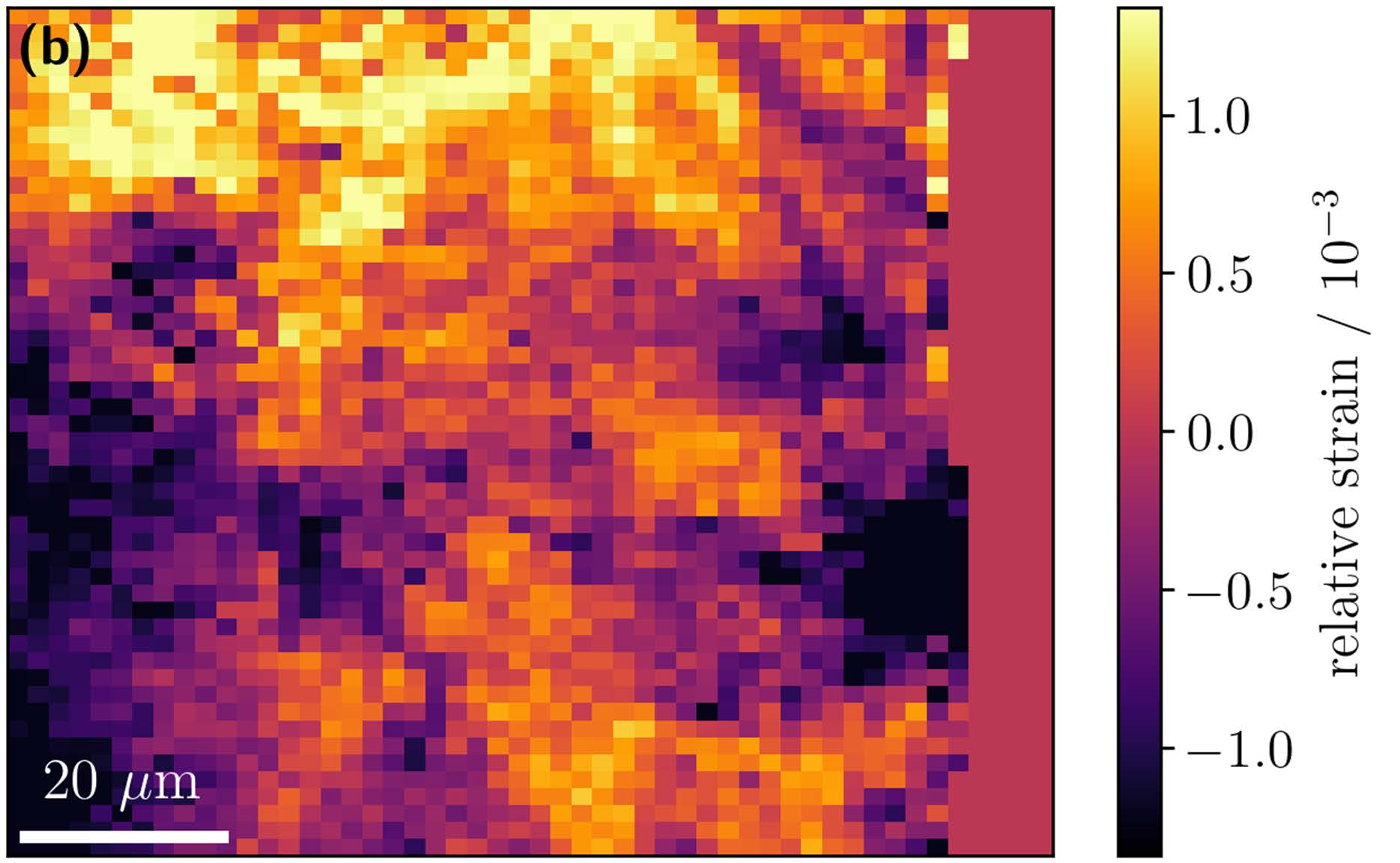}\\
\caption{\label{fig:data analysis_2}Relative strain maps in the (200) direction of two martensitic steel samples with a shot peening Almen intensity of 0.16~mm (a) and 0.3~mm (b). Images share the same colormap.}
\end{figure}

Fig.~\ref{fig:data analysis_2} compares the local relative strain in the [200] direction of two martensitic steel samples, which were treated with different shot peening intensities as mentioned above. Please note that although relative strains are displayed, the images are rooted in identical yet unknown absolute values, which allows for the utilization of the same colormap. As expected~\cite{Eleiche2001} shot peening induces a compressive strain in the steel samples that scales with shot peening intensity. Here, we measured an average compressive strain inducement of $\Delta \epsilon_{200} = 8\cdot 10^{-4}$. At the same time we observed that increased shot peening intensity also produces significant smaller micro-structures. This visual impression is confirmed by the autocorrelation lengths of 37~$\mu$m for a shot peening intensity of 0.16~mm (Fig.~\ref{fig:data analysis_2}a) and of 12~$\mu$m for 0.3~mm (Fig.~\ref{fig:data analysis_2}b), respectively. Out of beam focus conditions for (Fig.~\ref{fig:data analysis_2}a) was excluded by the fact that the goniometer was located on a kinematic mount and the longitudinal spot sizes of a few millimeters. Further, average peak counts in the analyzed reflections were 40,000 for (Fig.~\ref{fig:data analysis_2}a) and 17,000 for (Fig.~\ref{fig:data analysis_2}b), respectively. 

Ultimately, these examples demonstrate the capability of $\mu$-HRXRD at P06 for micro-structural characterization for highly absorbing materials. For example, by taking advantage of the additional rotational degrees of freedom ($\Psi$ \& $\Phi$ in Fig.~\ref{fig:setup}b) the full surface stress tensor of martensitic steel samples can be determined locally with micrometer spatial resolutions.

As an example for crystal structures providing well-defined single Bragg peaks we used a thin-film solar cell sample. For these material systems, the combination of a micro-diffraction goniometer with micro-XRF at high X-ray energies is compelling as it allows for a unique characterization of emerging photovoltaic materials: on one hand, the accurate analysis of the elemental distribution in the compound absorber layer requires the excitation of K-shell electrons. On the other hand, elemental inhomogeneities in the poly-crystalline semiconductors have been associated with strain-field variations and local solar-cell under-performance \cite{ulvestad-2019-jsr,strain:photovoltaic,correa-baena-2019-science}. 

Here, we demonstrate an application example of a solar cell with a (Ag,Cu)(In,Ga)Se$_2$ (ACIGS) absorber, where the local ratio of Ag/Cu \cite{aboufadl-2021-acsami} and In/Ga \cite{west-2017-nanoen} that occupy the same lattice points is critical for the device performance. Moreover, the 50~nm thick CdS layer forming the p-n junction together with the adjacent ACIGS and ZnO layers has been lacking spatial characterization in operational devices as the lateral distribution of Cd and In are only accessible with high-energy X-rays. The corresponding K absorption edges ($E(\mathrm{Cd}) = 27\,\mathrm{keV}$ and $E(\mathrm{In})= 28.0\,\mathrm{keV}$) are well accessible with the described setup.

The investigated solar cell was provided by Empa (Switzerland) and synthesized on soda-lime glass. The ACIGS absorber layer, which is of main interest, was grown on top of the Mo back electrode. Alkali metals including Na, K, or Rb were added in a post-deposition treatment for the passivation of defects at grain boundaries. The front electrode consists of a CdS layer in contact with the absorber and an optically transparent ZnO absorber top electrode.  For details of the synthesis conditions and sample properties, we refer to the description given by S.-C.~Yang, et al.~\cite{yang-2021-srrl}.

\begin{figure}
    \centering
    \centering
    \captionsetup{justification=raggedright,singlelinecheck = false}
    \includegraphics[width=0.48\textwidth]{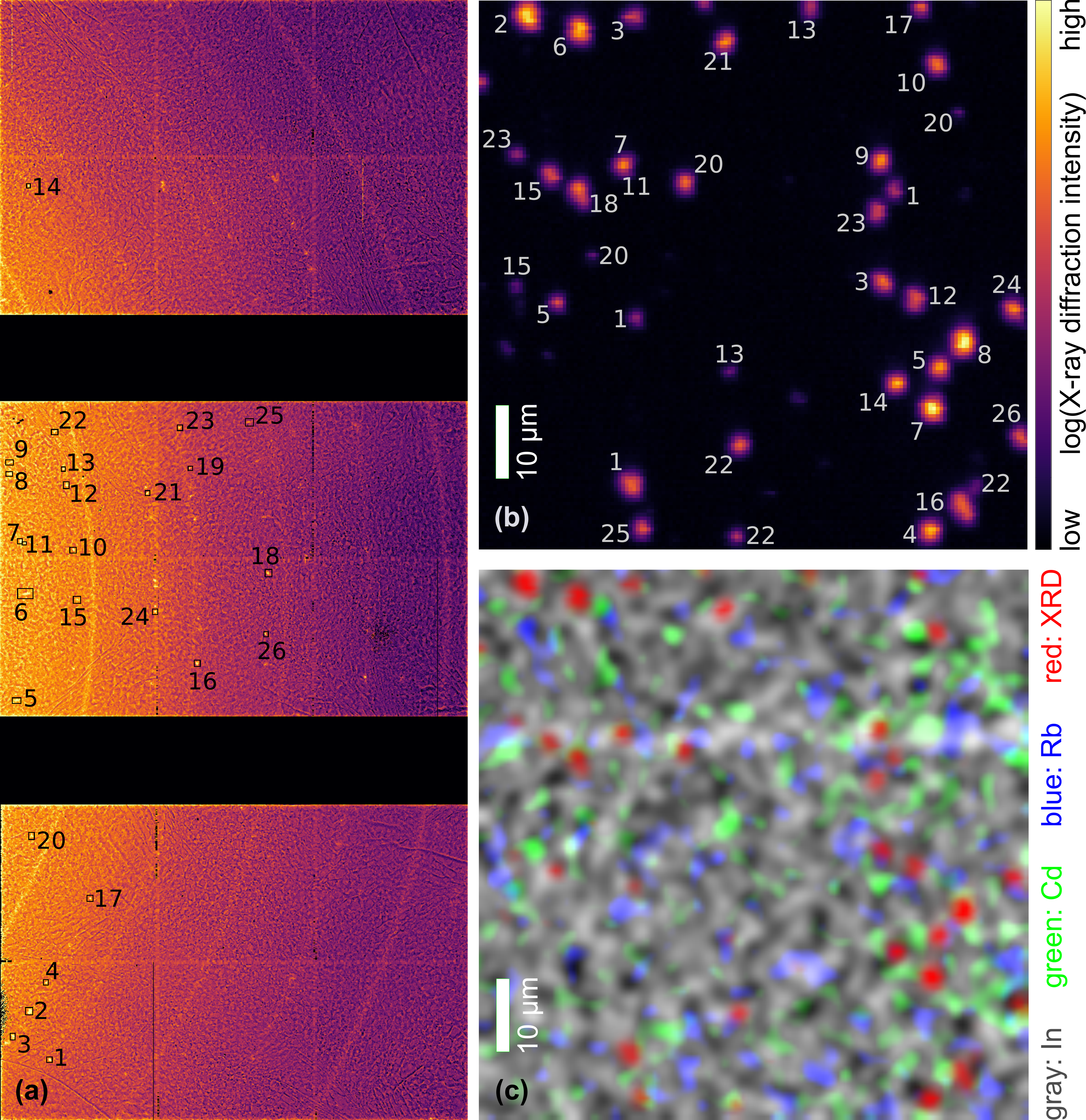}
    \caption{(a) X-ray diffraction detector image summed over the mapped area of a solar cell with a polycrystalline ACIGS absorber layer. The rectangles with adjacent numbers denote diffraction peaks from individual ACIGS crystallites. (b) Real-space map of the X-ray diffraction intensity with the crystallites attributed to their position in the reciprocal-space image shown in panel (a). For example, locations in (b) indicated by '1' provide the Bragg peak in (a) indicated by the same number. (c) Combined map of the X-ray diffraction intensity (red) with the elemental distributions of In (gray), Cd (green), and Rb (blue).}
    \label{fig:solar_cell}
\end{figure}

For the simultaneous assessment of structure and composition employing X-ray diffraction and fluorescence, we mapped a $({150\times 150)\,\mu}$m$^2$ large area of an ACIGS solar cell in plan view with a step size of $(1\times 1)\,\mu$m$^2$. In first approximation, the crystallites in the ACIGS material are randomly distributed and have a maximum size of few micrometers. Accordingly, the Bragg condition is only sporadically fulfilled, which enables the unambiguous distinction of individual crystallites in real space that would form powder rings in classical diffraction measurements with larger beam size.

The diffraction-detector images, integrated over the entire real-space map, are shown in Fig.~\ref{fig:solar_cell}(a). The peaks that could unambiguously be attributed to the ACIGS material are highlighted by black rectangles and enumerated for their localization in the real-space map shown in Fig.~\ref{fig:solar_cell}(b). The compositional variations within the ACIGS material and the tetragonal lattice lead to overlapping $2\theta$-ranges from distinct lattice planes and pose a particular challenge for the conversion of the lattice spacing into strain; the detailed discussion of this aspect is beyond the scope of this article and will be addressed in a dedicated article about correlative nano-diffraction results. 

The X-ray fluorescence data was fitted with PyMca \cite{heginbotham-2017-arch} and the resulting spatial distribution of In, Cd, and Rb are shown in Fig.~\ref{fig:solar_cell} along with the XRD peaks from ACIGS. To our knowledge, these are the first XRF maps of an ACIGS solar cell whose key elements were all assessed above their K absorption edge, which leads to significantly less self-absorption artifacts and more accurate quantification compared to assessment of the L-line X-ray fluorescence that strongly overlap.\cite{nietzold-2018-jove, ziska-2020-pvsc} The map shows an inhomogeneous distribution of In that arises mainly from topological variations: crevices and voids, predominantly at grain boundaries, are abundant in these solar cells \cite{avancini-2018-stam}. While the CdS layer--the only layer with Cd--is supposed to be homogeneously distributed laterally for optimum electrical properties, these measurements clearly unveil inhomogeneities and suggest that the chemical-bath deposition of CdS leads to partial filling of crevices and voids. Finally, the map shows that Rb is anti-correlated with In as expected, which is in agreement with earlier measurements~\cite{schoppe-2017-nanoen,plass-2020-acsami,ossig-2022-prep}. 

As the setup is readily compatible with scanning the rocking curve of individual ACIGS grains, their corresponding strain can be determined. This will allow us to correlate local elemental composition with micro-structural properties of ACIGS crystallites that fulfilled the Bragg condition during a scan. The number of the latter can be increased by additional scans with the sample rotated by $\Phi$ (see Fig.~\ref{fig:setup}b).

\section{Conclusions and Outlook}

In conclusion, we have demonstrated $\mu$-HRXRD at a photon energy of 35~keV, which allows to acquire the X-ray diffraction signal with micrometer spatial resolution from highly absorbing samples. We showed the feasibility of this approach with two different types of samples: martensitic steel as a powder-like and a thin film solar cell as a single crystal example.

For the martensite samples we were able to image boundaries of prior austenite grains as well as the lamella-like structure of the martensite phase. The observation of induced compressive strain with different shot peening intensities validated our method. By taking advantage of the additional rotational degrees of freedom provided by the goniometer, we will demonstrate that the full surface stress tensor can be determined by the proposed setup. This will enable us to study the impact of microscopic residual stress on short crack propagation in spring steels. In turn, this will provide a computational approach to the fracture-mechanical proof of fatigue strength in spring steels, which will be much faster and more efficient for structural integrity assessments compared to time consuming fatigue testing~\cite{wildeis2021}.

For thin-film ACIGS solar cells, we demonstrated the simultaneous assessment of XRD enabled characterization of the micro-structure with the elemental distribution of high-Z materials via XRF. The readily available extension to rocking curve imaging via 3D scans opens the possibility to correlate elemental distribution with local strain shedding light on the mechanisms of performance reduction. In addition, the K absorption edges of I, $E(\mathrm{I})  = 33.2$~keV, and Cs, $E(\mathrm{Cs})= 36.0$~keV) of perovskite solar cells are also accessible by increasing the photon energy to 42~keV.

Future improvements of the setup will include scan speed and focus size. Here, total scan times were limited not by the XRD signal (several 10,000 counts per scan point and diffraction peak) but by the overhead of utilised stepping positioners, which was kept to around 100\%. Thus, a reduction of scan times by at least a factor of 10 is readily feasible for "fly scan" capable positioners. Further, we have already demonstrated that focal spot sizes of down to 110~nm are achievable by utilizing phase plates for aberration corrections at 15 keV~\cite{Ossig2021} as well as 35~keV~\cite{Falkenberg:2020}. Thus, the setup will provide users from the materials science community the means to determine micro- and nano-structures as well as the elemental composition even in highly absorbing samples.

\section*{Acknowledgements}

We would like to thank Shi-Chi Yang, Romain Carron, and Ayodhya N. Tiwari (Empa) for providing a solar-cell sample and Patrik Wiljes and Stephan Botta (DESY) for experimental support. We acknowledge DESY (Hamburg, Germany), a member of the Helmholtz Association HGF, for the provision of experimental facilities. This research was supported in part through the Maxwell computational resources operated at DESY.

\section*{References}
\bibliography{HRXRD.bib}

\end{document}